\documentclass[preprint,superscriptaddress, amsmath,amssymb,aps,floatfix]{revtex4-1}

\usepackage[english]{babel}
\usepackage[utf8]{inputenc}
\usepackage{graphicx}
\usepackage{todonotes}
\usepackage{bm}
\usepackage{xcolor}
\usepackage{hyperref}
\usepackage{amsthm}

\newtheorem{theorem}{Theorem}

\newcommand{\EE}{\mathcal{E}}

\newcommand{\LL}{\mathcal{L}}

\begin{document}

\title{Discontinuous transition to loop formation in optimal supply networks}

\author{Franz Kaiser}%
 \email{f.kaiser@fz-juelich.de}
  \affiliation{Forschungszentrum J\"ulich, Institute for Energy and Climate Research (IEK-STE), 52428 J\"ulich, Germany}
    \affiliation{Institute for Theoretical Physics, University of Cologne, K\"oln, 50937, Germany}
    \author{Henrik Ronellenfitsch}
\affiliation{Department of Mathematics, Massachusetts Institute of Technology, Cambridge, MA 02139, U.S.A.}
\affiliation{Physics Department, Williams College, 33 Lab Campus Drive, Williamstown, MA 01267, U.S.A.}
\author{Dirk Witthaut}%
 \email{d.witthaut@fz-juelich.de}
  \affiliation{Forschungszentrum J\"ulich, Institute for Energy and Climate Research (IEK-STE), 52428 J\"ulich, Germany}
    \affiliation{Institute for Theoretical Physics, University of Cologne, K\"oln, 50937, Germany}    

\begin{abstract}
The structure and design of optimal supply networks is an important topic in complex networks research. A fundamental trait of natural and man-made networks is the emergence of loops and the trade-off governing their formation: adding redundant edges to supply networks is costly, yet beneficial for resilience. Loops typically form when costs for new edges are small or inputs uncertain. Here, we shed further light on the transition to loop formation. We demonstrate that loops emerge discontinuously when decreasing the costs for new edges for both an edge-damage model and a fluctuating sink model. Mathematically, new loops are shown to form through a saddle-node bifurcation. Our analysis allows to heuristically predict the location and cost where the first loop emerges. Finally, we unveil an intimate relationship among betweenness measures and optimal tree networks. Our results can be used to understand the evolution of loop formation in real-world biological networks.
\end{abstract}

\maketitle

\clearpage
\newpage
\section{Introduction}
The reliable function of supply networks is essential for biological as well as technical systems. Leaf venation networks supply plant leaves with water and nutrients~\cite{Sack2013} and vascular systems supply vertebrates with oxygen and nutrients~\cite{Pittmann2011}. On the other hand, society relies on man-made supply networks such as power grids~\cite{Wood14} or hydraulic networks~\cite{Hwan96}. Finally, networks that formed over time such as drainage basins show a similar structure~\cite{rodriguez2001fractal}. Understanding the design principles of such networks is a central challenge in network science \cite{havlin_challenges_2012}.

The evolution or construction of supply and transportation networks is essentially determined by the trade-off between cost and resilience~\cite{Ronellenfitsch2017phenotypes,ganin_resilience_2017,avena-koenigsberger_path_2017}. Cost limits the number of connections in the network, as resources are generally scarce. Resilience requires additional connections to cope with damages or perturbations. Many actual networks contain loops to establish a certain level of topological resilience, hence they stay connected and operational even if some elements fail~\cite{katifori_quantifying_2012}. The interplay of topology and resilience is analysed in various disciplines including traffic networks \cite{ganin_resilience_2017}, communication networks \cite{sterbenz2010resilience} or dynamical networks \cite{gao2016universal}. Finally, a variety of results on structural resilience, that is the ability of a network to remain connected when a fraction of nodes or links fails, have been obtained in network science \cite{shargel2003optimization,paul2004optimization}. 

\begin{figure}[tb]
    \begin{center}
        \includegraphics[width=0.5\columnwidth]{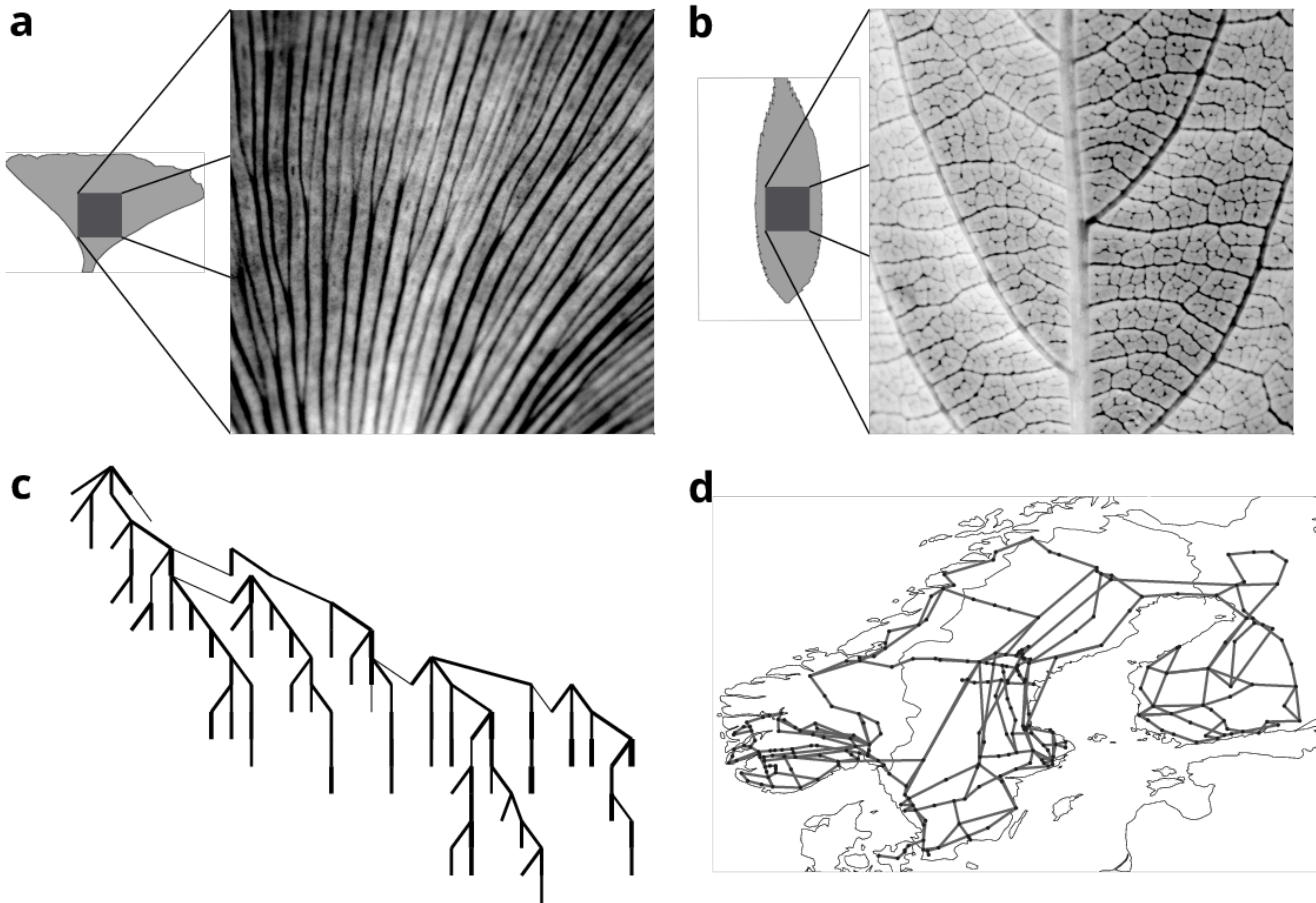}%
    \end{center}
    \caption{\textbf{Loopy and non-loopy real-world supply networks.} \textbf{a} The leaves of \emph{Ginkgo biloba} and \textbf{c} the distribution grid IEEE123 form loopless supply networks. \textbf{b} The venation network of \emph{Prunus serrulata} and \textbf{d} the Scandinavian power grid on the transmission level form loopy supply networks. Leaf venation networks extracted from photographs, distribution grid taken from Ref.~\cite{Kersting_2001} and transmission grid topology extracted from the open power system model PyPSA-Eur~\cite{horsch_2018}.
     }
    \label{fig:real_world_networks}
\end{figure}

In this article, we focus on linear flow networks modelling power grids, hydraulic networks or vascular networks~\cite{katifori2010damage,Hwan96,Wood14}. Different structural patterns are observed in nature, consisting of both networks with and without loops. For instance, leaf venation networks are loopy in general, except for a few old species such as \emph{Ginkgo}. In electric power systems, large-scale transmission grids are strongly meshed, while local distribution grids are topological trees (see Figure~\ref{fig:real_world_networks}). Optimal network structures balancing costs and resilience have been analysed via extensive numerical simulations in the setting where a single source supplies the remaining network, such as in plant leaves~\cite{corson2010fluctuations,katifori2010damage,bohn_structure_2007}. The optimal structure does not contain any loops if connections are reliable and perturbations are weak, for instance in distribution grids. Loops come into being when sources or sinks fluctuate strongly or connections are subject to damages, such as in transmission grids or newer leaf species. While some work has been done in the context of networks optimising transport time~\cite{Kirkegaard2020}, the exact mechanism of loop formation in minimal-dissipation networks is still not fully understood. 

Here, we analyse the transition to loop formation on a theoretical basis and derive several analytical results. We consider optimal network structures in the sense that function is optimised while costs are constrained or vice versa. Two aspects of resilience are studied in detail -- damage to edges and fluctuations of supply and demand. In particular, we investigate the optimal structure as a function of the severity of damage and the strength of fluctuations. In contrast to prior work we focus on the occurrence of the very first loop, which enables an analytical approach to loop formation and yields several rigorous results. We first establish this approach for an elementary sample network and then generalise it to networks of arbitrary size and compare analytic predictions and numerical results.

In particular, we demonstrate that the transition to loop formation is generally {discontinuous} in the sense that optimal edge-capacities jump discontinuously when fluctuations increase or costs decrease. Loopy network structures emerge as new local minima of the dissipation function that form via a saddle-node bifurcation, and not via a bifurcation of an already existing minimum. Hence, a large number of local minima may exist simultaneously and we establish a purely topological expression based on the edge betweenness to understand their structure. As a direct application of our analysis we derive a simple criterion to predict the location of the first loop in the transition from a tree network.\\ \\
\section{Results}
\textbf{Modelling supply networks.}
We consider a simple supply network model which was previously used to study loop formation in generic distribution networks~\cite{corson2010fluctuations,katifori2010damage}. Mathematically, the supply network is constructed from a graph $G$ with node set $V$ and edge set $E$. At each node $n \in V$ there is an in- or outflow with a strength $P_n$, where $P_n>0$ denotes a source and $P_n<0$ a sink. The in- and outflows may either represent individual supply nodes or allocated demands associated with the node~\cite{farina_using_2014}. An edge in the network is either labelled by its index $e \in E$ or by its terminal nodes $e=(n,m)$ which we use interchangeably. For each edge we fix an orientation which is encoded in the node-edge incidence matrix $I \in \mathbb{R}^{|V| \times |E|}$ with elements
\begin{equation}
   I_{n,e} = \left\{
   \begin{array}{r l}
      1 & \; \mbox{if edge $e$ starts at node $n$},  \\
      - 1 & \; \mbox{if edge $e$ ends at node $n$},  \\
      0     & \; \mbox{otherwise}.
  \end{array} \right.
  \label{eq:def-nodeedge}    
\end{equation}
Each edge is assigned a capacity $k_e \in \mathbb{R}_+$ and a flow whose strength or value is denoted as $F_e \in \mathbb{R}$. Fixing the orientation of an edge $e = (n,m)$  means that $F_e > 0$ describes a flow from node $n$ to node $m$ and $F_e < 0$ describes a flow from node $m$ to node $n$. The flows satisfy the continuity equation or Kirchhoff's current law (KCL) at every node of the network,
\begin{equation}
    \sum_{e \in E} I_{n,e} F_e = P_n, \qquad
    \forall \, n \in V.
    \label{eq:KCL}
\end{equation}
In addition to that, we assign a potential $\theta_n$ to each node in the network. In terms of physical quantities, this potential $\theta_n \in \mathbb{R}$ can represent the pressure at the nodes of a hydraulic network, the voltage in DC resistor networks or the nodal voltage phase angle in linearised AC power grids~\cite{Hwan96,diaz2016,katifori2010damage,strake_non-local_2019}. For these systems, the flow on a link $e =(n,m)$ scales linearly with the potential drop $\theta_n-\theta_m$ along the link and can be calculated as
\begin{equation}
\begin{aligned}
F_e = k_e (\theta_n-\theta_m).\label{eq:kirchhoff}
    \end{aligned}
\end{equation}
Together with the continuity equation~\eqref{eq:KCL}, this linear set of equations determines the values of the potentials $\theta_n$ up to a global constant. The resulting flows automatically satisfy Kirchhoff's voltage law (KVL) which states that the flow around any closed loop expressed in terms of the edges $C=\{e_{c_1},e_{c_2},...,e_{c_{\text{max}}}\}$ vanishes~\cite[pp.40]{bollobas_modern_1998}
\begin{equation}\begin{aligned}
    \sum_{e\in C} z_e F_e = 0.\label{eq:KVL}
\end{aligned}\end{equation}
Here, the factor $z_e\in\{-1,1\}$ is used to keep track of the orientation of an edge $e$ with respect to the orientation of the edge in the loop $C$, i.e. 
\begin{equation*}
z_e = 
\left\{
   \begin{array}{r l}
   1 & \; \mbox{ if edge } e=(e_1,e_2) \mbox{ is oriented from } e_1 \mbox{ to } e_2,\\
   -1 & \; \mbox{ if edge } e=(e_1,e_2) \mbox{ is oriented from } e_2 \mbox{ to } e_1.
   \end{array} \right.
\end{equation*}

\noindent\textbf{Optimising supply networks: minimum dissipation topologies.}\\
We now illustrate how to determine the {optimal} supply network that is described by the above the set of equations. To this end, we want to find the edge capacities that determine the network structure that is optimal for performing a given task. Throughout this manuscript, we call the network structure {optimal} if the edge capacities are such that the overall network dissipation is minimised, as suggested for example in Refs.~\cite{corson2010fluctuations,katifori2010damage,bohn_structure_2007}.  The network dissipation may be calculated as
\begin{equation}\begin{aligned}
    D = \sum_{e \in E} \frac{F_e^2}{k_e} \, .\label{eq:dissipation}
\end{aligned}\end{equation}
In addition to that, we assume that the resources to build the network are limited. This resource constraint takes the form
\begin{equation}\begin{aligned}
    \sum_{e \in E} k_e^{\gamma} = K^\gamma,
    \label{eq:capacity-constraint}
\end{aligned}\end{equation}
where the cost parameter $\gamma>0$ depends on the type of problem under consideration. For instance, assuming Poiseuille flow through cylindrical
pipes of fixed length and radius $R_e$, $k_e \sim R_e^4$, such that $\gamma=1/2$ fixes 
total fluid volume and $\gamma=1/4$ fixes total pipe mass~\cite{Durand2007,corson2010fluctuations,katifori2010damage,banavar_topology_2000,bohn_structure_2007}.
The parameter $K$ corresponds to the overall available budget. Note that different definitions of optimal networks arise in other applications, e.g. in hydraulic engineering where typically the cost is minimised while the dissipation is constrained~\cite{creaco_fast_2012}. In Supplementary Figure 5 and Supplementary Note 5, we demonstrate that the same kind of discontinuous transition is observed when extending our analysis to this setup.

In general, it is neither useful nor meaningful to allow arbitrary connections between the nodes. Geometric constraints apply to a variety of networks. For instance, leaf vascular networks or river basins are naturally planar. To take into account such constraints and keep the problem feasible one typically fixes a set of {potential edges} $\EE$ such that $E \subseteq \EE$. These edges are often taken from a square grid~\cite{corson2010fluctuations}, a triangular grid~\cite{katifori2010damage}, or various types of disordered tessellations~\cite{ronellenfitsch_global_2016,Ronellenfitsch2017phenotypes}. Note that while planarity of the network described by the set of potential edges $\EE$ simplifies the theoretical analysis, our results are not limited to planar networks as we demonstrate for a simple, non-planar network in Supplementary Figure 6.

We focus on two different models here: a model with fluctuating sources and sinks and a model of stochastic damage to the edges. Both models can be thought of as quantifying network resilience: We call a network resilient if it is able to function properly under the uncertainities induced by edge damage or fluctuating inputs. For both models, our main question will be the following: Under which conditions does the optimal network structure contain loops and how do these loops emerge?

\noindent{Fluctuating sink model.}
First, we introduce the fluctuating sink model. In this model, we include fluctuations by treating the $P_n$ as random variables. For each random realisation, the sources and sinks are balanced, i.e. they sum to zero,
\begin{equation}\begin{aligned}
    \sum_{n \in V} P_n = 0.\label{eq:balanced}
\end{aligned}\end{equation}
Network structures are then optimised to have a minimum {average} dissipation
\begin{equation}\begin{aligned}
    \langle D\rangle = \sum_{e \in E} \frac{\langle F_e^2 \rangle}{k_e} , 
    \label{eq:diss}
\end{aligned}\end{equation}
for a given set of resources. Here, the brackets $\langle \cdot \rangle$ denote the expected value taken over all realisations of the random variables $P_n$. Note that the fluctuations affect only the flows directly by virtue of Eq.~\eqref{eq:kirchhoff}, whereas the network topology is assumed to be fixed by the construction of the network such that the average is taken over the squared flows only. Equation~\eqref{eq:diss} can be minimised analytically with respect to the $k_e$, where the resource constraint is taken into account via the method of Lagrange multipliers. Calculating the optimal edge capacities by extremising the Lagrange function yields~\cite{corson2010fluctuations} (see Supplementary Note 1)%
\begin{equation}\begin{aligned}
    k_e = \frac{
    (\langle F_e^2 \rangle)^{\frac{1}{1+\gamma}} 
    }{
    \left[ \sum_{a\in E} (\langle F_a^2 \rangle)^{\frac{\gamma}{1+\gamma}}  \right]^{1/\gamma}
    } 
    K.
    \label{eq:ke_corson}
\end{aligned}\end{equation}
This expression depends on the second moments of the flows $\langle F_e^2\rangle$, which in turn depend on the capacities $k_e$. Hence, Eq.~\eqref{eq:ke_corson} can be interpreted as a self-consistency condition which has to be solved together with Eq.~\eqref{eq:kirchhoff}. 

\begin{figure*}[tb]
    \begin{center}
        \includegraphics[width=1.0\textwidth]{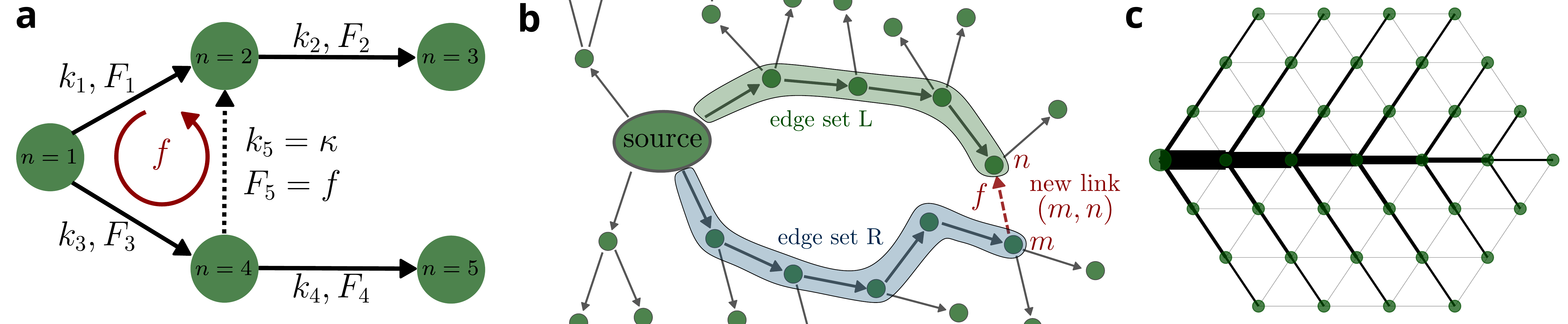}%
    \end{center}
    \caption{\textbf{Graph setup to analyse the transition from tree networks to loopy networks.}
     \textbf{a} Elementary network to study spontaneous loop formation in optimum supply networks. The network consists of five nodes (green circles) where node $n=1$ has an inflow of four, $P_1=4$, and all other nodes have an outflow of unity. These in and outputs determine the flows $F_i,i\in\{1,2,3,4,5\}$ along the links with capacities $k_i$. The optimum topology for this set-up is a tree network. If the in- and outputs are fluctuating, an additional edge (dotted arrow) may be beneficial to reduce the average dissipation. This edge introduces a new degree of freedom expressed as a cycle flow $f$. \textbf{b} For a larger network, we generalise this setup as follows: we start from a tree network and then consider the impact of a new edge at an arbitrary position $(n,m)$ (dotted, red arrow). We then collect the edge sets $L$ (shaded green) and $R$ (shaded blue) along the shortest path from the source to the newly formed edge. This edge induces a cycle flow $f$. \textbf{c} A network formed from a triangular grid with set of potential edges $\EE$ coloured in grey which we will analyse throughout the manuscript. Realised edges (black) correspond to a global minimum of the dissipation for the fluctuating sink model where a single, fluctuating source (large circle) supplies the remaining network.
     }
    \label{fig:5node_network}
\end{figure*}

\noindent{Edge-damage model.} A second class of dissipation-optimised networks that is relevant to biology and engineering seeks to find optimal networks subject to damage. For instance, leaf vasculature might be attacked by a herbivorous insect, or a power grid might lose a power line due to an outage. In the following, we generalise the broken-bond model considered in Ref.~\cite{katifori2010damage} by allowing partial damage to the network capacities instead of a complete removal of edges. 

 In this edge-damage model, the sources and sinks are still balanced, but do not fluctuate stochastically. Instead, we assume that all nodes but one are sinks with $P_{j>1} = - \bar P$ supplied by a single node with $P_1 = (N-1)\bar P$, where $N$ is the number of nodes. 

To model partial damage of edge $l$, we modify the edge capacities according to
\begin{equation}\begin{aligned}
    k_e \, \rightarrow \, (1-\Delta_e^{(l)}) k_e
\end{aligned}\end{equation}
with the damage fraction 
\begin{equation}\begin{aligned}
    \Delta_{e}^{(l)} = 
    \left\{
   \begin{array}{l l}
    0 & \; \mbox{ if  $e \neq l$},\\
    \Delta\in (0,1] & \; \mbox{ if  $e=l$}.
    \end{array}\right.
\end{aligned}\end{equation}
Thus, a damage parameter $\Delta=1$ corresponds to a complete removal of the damaged edge. We now define the average over all possible damage scenarios. Specifically, if $g(k_e)$ is some function of the capacities $k_e$, we define
\begin{equation}\begin{aligned}
    \langle g(k_e) \rangle' = \frac{1}{|E|}\sum_{l=1}^{|E|} g(\Delta_{e}^{(l)} k_e),
\end{aligned}\end{equation}
where $|E|$ is the number of edges in the network. Here and in the following we use the notation $\langle \cdot \rangle'$ to distinguish the average over damage scenarios from the average over fluctuating sources and sinks.

As before, the central objective is to minimise the average dissipation of the network,
\begin{equation}\begin{aligned}
    \langle D \rangle =  \sum_e \left\langle \frac{F_e^2}{k_e}, \right\rangle',
\end{aligned}\end{equation}
taken over all possible damaged edges under the resource constraint Eq.~\eqref{eq:capacity-constraint}.

We now proceed to study loop formation in the two models outlined above in detail. \\[1em]

\normalsize \textbf{Discontinuous transition to loop formation in small network.} As an illustrative example, let us consider an elementary network as sketched in Figure \ref{fig:5node_network} a and analyse the transition to loop formation in both, the fluctuating sink model and the edge-damage model. \\
{Discontinuous transition in fluctuating sink model.} The network consists of four variable sinks at nodes $2,3,4,5$ (circles) that are modelled as iid Gaussian random variables $P_{2,3,4,5} \sim \mathcal{N}(\mu,\sigma)$ and four edges (arrows) connecting them with capacities $k_i$ and flows $F_i,i\in\{1,2,3,4\}$. A fifth, potential edge is shown as dotted arrow. If it exists, it carries flow $\tilde{F}_5$ and has capacity $k_5 = \kappa$ (Fig.~\ref{fig:5node_network} a). The central question we will study for this setup is the following: When is the optimal network tree-like ($\kappa=0$) and when is it loopy ($\kappa>0$) -- and how does $\kappa$ behave at the transition point?  

We first consider the case where the loop is not present, i.e. $\kappa=0$. In this case, the network is a tree and we can calculate the second moments $\langle F_i^2 \rangle,~i\in\{1,2,3,4\}$ explicitly in terms of the capacities: they are determined by the statistics of the source and the sinks by virtue of the continuity equation (\ref{eq:KCL}). Using the optimal capacities for a tree network (\ref{eq:capacity-constraint}), we obtain an explicit equation for the optimal dissipation $\langle D_{\rm tree} \rangle$ that only depends on the statistics of the sinks (see Supplementary Note 4)
\begin{equation}\begin{aligned}
    \langle D_{\rm tree} \rangle
       &= \frac{\left[2(\sigma^2+\mu^2)^{\frac{\gamma}{\gamma +1}}+2(2\sigma^2+4\mu^2)^{\frac{\gamma}{\gamma +1}}\right]^{(\gamma +1)/\gamma}}{K}.
\end{aligned}\end{equation}

How does this result change if we allow to close the loop as illustrated in Figure~\ref{fig:5node_network} a, i.e., if we include the corresponding edge in the set of potential edges $\EE$?

\begin{figure*}[tb]
    \begin{center}
      \includegraphics[width=0.9\textwidth]{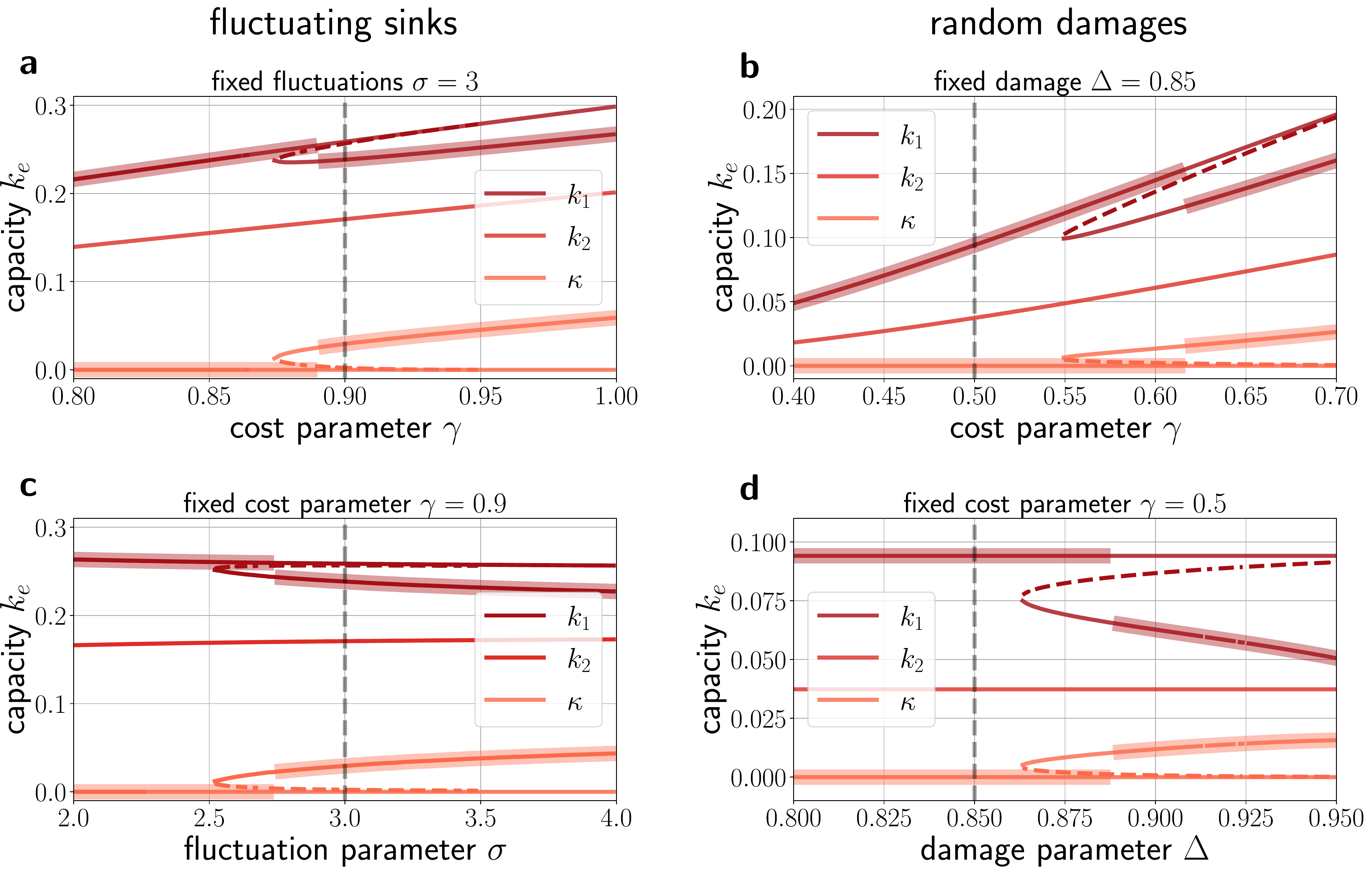}
    \end{center}
    \caption{\textbf{Discontinuous transition in dissipation minimum appears throughout models and parameters.} Capacities at the global minimum (thick lines) show a discontinuity for different models when analysing the topology shown in Figure~\ref{fig:5node_network} a. \textbf{a,c} We analyse the edge capacities $k_e$ at the local minima (straight lines) and saddle (dotted lines) for varying cost parameter $\gamma$ (a) and varying fluctuation parameter $\sigma$ (c) for the fluctuating sink model with fluctuation mean $\mu=-1$ and total capacity $K=1$. For both parameters, the capacity at the loop $\kappa$ (light orange) undergoes a saddle node bifurcation which causes a discontinuous transition in the global minima (thick lines) from non-loopy to loopy networks. \textbf{b,d} An analogous saddle-node bifurcation in the capacities $k_e$ may be observed in the generalised damaged bond model in terms of both the cost parameter (b) and the damage parameter (d). For all four plots, dotted black lines denote the matching values in the other plot. }
    \label{fig:saddle_node_bifurcation}
\end{figure*}
Let us assume that the loop carries a flow $\tilde F_5$ and has a non-zero capacity $k_5 = \kappa>0$. In the following, we denote the flows and capacities in the loopy network with a tilde. In the presence of a loop, we can no longer determine the flows using the continuity equation~\eqref{eq:KCL} alone. Instead, we have an additional degree of freedom: a cycle flow $f$ around the newly formed edge such that $\tilde F_1 = F_1 – f$, $\tilde F_3 = F_3 + f$ and $\tilde F_5 = f$. The strength of the cycle flow can be determined using the KVL (\ref{eq:KVL})
\begin{equation}\begin{aligned}
    & \frac{f}{\kappa} + \frac{\tilde F_3}{\tilde k_1} 
    - \frac{\tilde F_1}{\tilde k_1} = 0.
\end{aligned}\end{equation}

This approach allows us to eliminate the dependence on the cycle flow strength $f$, and we can evaluate the dissipation $\langle D_{\rm loopy}\rangle$ of the loopy network by inserting the result into Eq.~(\ref{eq:diss}) (see Supplementary Note 4). %
The new expression for the dissipation no longer contains the flows explicitly, which considerably simplifies finding the optimal topology: we no longer have to take care of the interdependence of flows and capacities, but can minimise $\langle D_{\rm loopy}\rangle$ in terms of only the capacities $\tilde{k}_i$.

We proceed to evaluate the optimal network structure fixing the mean of fluctuations to $\mu=-1$ and the resource constraint to $K=1$. To examine the effect of the two remaining parameters
separately, we analyse the transition to loop formation for $\gamma = 0.9$ fixed while varying $\sigma$ and for $\sigma=3$ fixed with varying $\gamma$. We then compute the dissipation $\langle D_{\rm loopy}\rangle$ as a function of the capacities $\kappa$ and $k_1$ and compare it to the dissipation $\langle D_{\rm tree}\rangle$ of the corresponding tree network. Note that the capacities in the optimum tree network are explicitly given by Eq.~\eqref{eq:ke_corson} such that $\langle D_{\rm tree}\rangle$ is fixed. For the loopy network we still need to determine the optimum structure, i.e. we compute the minima of $\langle D_{\rm loopy}\rangle$ as a function of $\kappa$ and $\tilde k_1$ recalling that $\tilde k_3=\tilde k_1$, $\tilde k_4=\tilde k_2$ and $\tilde{k}_2$ is then fixed by the resource constraint Eq.~\eqref{eq:capacity-constraint}. 

For both varying fluctuations $\sigma$ and varying costs $\gamma$, we find that the transition to loop formation is discontinuous: the loop starts to form with a non-zero capacity $\kappa$ when analysing the globally optimal network structure (Fig.~\ref{fig:saddle_node_bifurcation} a,c, thick, orange line). Analogously, the capacity $k_1$ bifurcates (red line). 

But what is the nature of this transition? In fact, we find that new minima emerge through a saddle-node bifurcation independently of the parameter we vary. Thus, new minima do not form from the existing tree minimum, but instead emerge elsewhere in the energy landscape. To support this claim, we plot the capacity at the saddle in Figure~\ref{fig:saddle_node_bifurcation} (dotted, coloured lines) and analyse the dissipation landscape close to the bifurcation point (see Supplementary Figure 3). %
Using these results, we can also map out the parameter region where loop formation is beneficial (see Supplementary Figure 2). In Supplementary Figure 7, we illustrate the nature of this transition for an even simpler network and find a closed-form solution for the region of the parameter space where loop formation is beneficial. %
\\ 
{Discontinuous transition in edge-damage model.}
We now turn to the edge-damage model and analyse the optimal topology again for the graph shown in Fig.~\ref{fig:5node_network} a. Most importantly, we find that the transition between a treelike and a loopy  optimal network is also discontinuous in the damage model in both the cost parameter $\gamma$ and the damage parameter $\Delta$, and new extrema appear again through saddle-node bifurcations (Fig.~\ref{fig:saddle_node_bifurcation} b,d).
This demonstrates that despite the fact that in the damage model, the optima follow a different scaling law from those in the fluctuation model~\cite{katifori2010damage}, the mechanism and type of the transition from tree-like to loopy optimum is generic.\\ 
\\\textbf{Discontinuous transition persists beyond the first loop.}
Whereas the transition to the {first} loop that forms is important in many real-world supply networks, such as the \textit{Gingko} leaf and the distribution grid shown in Figure~\ref{fig:real_world_networks}, other networks display several loops, such that their formation beyond the first loop becomes important. In particular, the tree has mainly a theoretical importance in many applications such as hydraulic networks where spanning trees in loopy networks play an important role in modelling and optimisation~\cite{elhay_reformulated_2014,creaco_comparison_2014,creaco_fast_2012,ciaponi_importance_2017}. Remarkably, we can demonstrate numerically that the discontinuous character of loop formation persists beyond the first loop.
\begin{figure*}[tb]
    \begin{center}
        \includegraphics[width=1.0\textwidth]{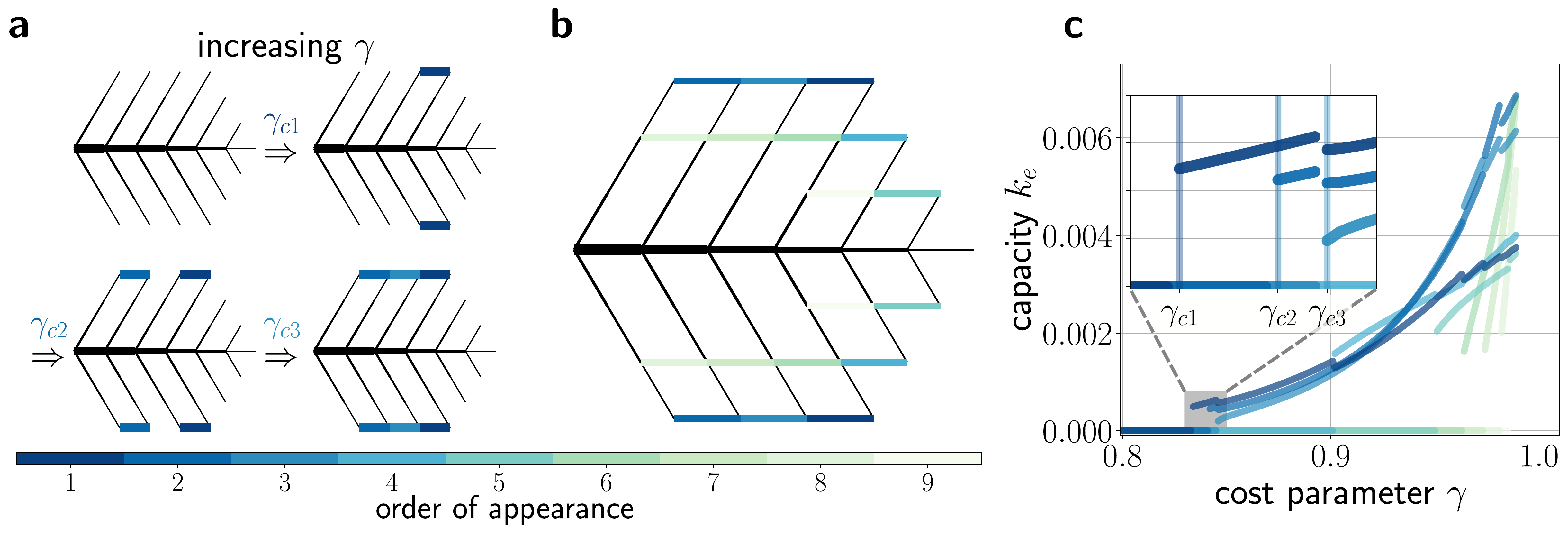}
    \end{center}
    \caption{\textbf{Discontinuous transition to loop formation persists beyond the first loop.} \textbf{a,b} We order the loops in a colour code according to their appearance with increasing cost parameter $\gamma$: the darker the edge colour, the earlier the edge appears. For the loop that appears as the $i$-th loop, we denote its critical cost parameter $\gamma_{ci}$ where the loop starts to become beneficial for the dissipation-optimised network. \textbf{c} The transition to loop formation is discontinuous beyond the first loop: loops appearing at higher values of $\gamma$ again appear with a non-zero capacity as shown in detail in the inset. Fluctuation strength is fixed to $\sigma = 0.5$ for all panels.}
    \label{fig:discontinuous_transition_larger_network}
\end{figure*}

 In Figure~\ref{fig:discontinuous_transition_larger_network}, we analyse this transition for the fluctuating sink model with cost parameter $\gamma = 0.5$ for a larger, globally optimal tree network which was formed from a set of potential edges $\EE$ corresponding to a triangular grid as shown in Figure~\ref{fig:5node_network} c. We map out the order in which new loops form (colour code) when decreasing the cost for new edges and slightly perturbing the previous network structure. All new loops emerge discontinuously with a non-zero capacity from an existing loopy network topology (Figure~\ref{fig:discontinuous_transition_larger_network} c). Note that in contrast to Figure~\ref{fig:saddle_node_bifurcation}, the optimal capacities are obtained here using an iterative approach for finding local minima of the dissipation that is due to Ref.~\cite{corson2010fluctuations} (see Methods). In a Supplementary Figure 8, we demonstrate that an analogous transition exists for varying fluctuation strength $\sigma$ and fixed cost parameter $\gamma$.\\ \\
\textbf{Identifying optimal trees for networks of arbitrary size.}
We now generalise our reasoning to larger networks with an arbitrary number of nodes $N$. For this analysis, we focus on the fluctuating sink model. Again we assume that all nodes $j=2,\ldots,N$ act as sinks with $P_j$ being random variables and that the source $j=1$ balances the sinks. We start from a tree network and analyse at what value of the cost parameter $\gamma$ it becomes beneficial to add a single edge thus closing a single loop. This setup is sketched in Figure~\ref{fig:5node_network} b. We first demonstrate how to calculate the dissipation in such a setting and then illustrate the procedure to minimise it.

In an arbitrary tree network, the flows do {not} depend on the link capacities but {only} on the topology of the network as illustrated in the last section. This is due to the fact that for each node $j=2,3,\ldots$ there is only one path from the respective node to the root $j=1$ of the tree. The flow $F_e$ on an edge $e$ is thus directly given by the KCL Eq.~\eqref{eq:KCL}. Here, we fix the orientation of the flows such that they point away from the source as illustrated in Figure~\ref{fig:5node_network} b. Therefore, flows in tree networks are always positive.

To express the flows $F_e$ in terms of the sources and sinks $P_j$, we introduce the tree matrix $\mathcal{T} \in \mathbb{R}^{|E| \times |E|}$ by
\begin{equation}
    \mathcal{T}_{e,j} = 
    \begin{cases}
    +1 & \parbox[t]{.3\textwidth}{if edge $e$ is on the path from node $j+1$ to the root $j=1$}, \\
    0  & \mbox{otherwise}.
    \end{cases}
    \label{eq:tree_matrix}
\end{equation}
This yields an explicit expression for the flows,
    \begin{equation}\begin{aligned}
        F_e &= -  \sum_{j=2}^N \mathcal{T}_{e,j-1} P_j \, . \label{eq:flows_treematrix}
    \end{aligned}\end{equation}
We can insert this result into the network dissipation (Eq.~\ref{eq:diss}), which yields
\begin{equation}\begin{aligned}
    \langle D_{\rm tree} \rangle
    &= \sum_{e \in T}  \sum_{i,j=2}^N 
       \mathcal{T} _{e,j-1} \mathcal{T}_{e,i-1} \langle P_i P_j \rangle k_e^{-1} \, ,
     \label{eq:D-tree}  
\end{aligned}\end{equation}
where $T=E(G)$ is the set of all edges in the tree, i.e. before the addition of a loop.\\ \\
\textbf{From trees to loopy networks: Optimising networks with a single loop.}
Remarkably, we can also find an explicit expression for the dissipation eliminating the flows in near-tree network by exploiting the KVL to eliminate the new degrees of freedom, similar to the strategy in the previous section. 

We consider a network that consists of a tree plus a single link $\ell = (m,n)$ with capacity $\kappa$ as sketched in Figure~\ref{fig:5node_network} b. The edges on the paths from nodes $n$ and $m$ to the root node are summarised in the edge sets $L$ and $R$ respectively, which we define as follows: Denote by $p(m)$ and $p(n)$ the set of all edges along the shortest path from the source node to the node $m$ and $n$, respectively, oriented in the direction pointing away from the source. Note that these paths are unique in a tree network. Then define the following sets:
\begin{equation}\begin{aligned}
    L &= p(n)\setminus (p(m)\cap p(n)),\\
    R &= p(m)\setminus (p(m)\cap p(n)),
\end{aligned}\end{equation}
such that the union of the edge set $L\cup R \cup \{(m,n)\}$ forms a cycle. As we will see in the following, this definition turns out to be useful when studying the dissipation in the presence of a single loop.

Due to the presence of the loop we have a new degree of freedom, the cycle flow strength $f$. According to the KCL Eq.~\eqref{eq:KCL}, the flows in the loopy network are given by
\begin{equation}\begin{aligned}
    \tilde F_e = \left\{
    \begin{array}{l l l}
      f & \; {\rm if} \;  e = (m,n) \\
      F_e + f & \; {\rm if} \;  e \in R(m) \\
      F_e - f & \; {\rm if} \;  e \in L(n) \\
      F_e     & \; {\rm otherwise }.
    \end{array} \right.
    \label{eq:change-of-F-new}
\end{aligned}\end{equation}
The value of $f$ is fixed via the KVL:
\begin{equation}\begin{aligned}
    & \sum_{e \in L} \frac{-F_e + f}{\tilde k_e}+
    \sum_{e \in R} \frac{F_e + f}{\tilde k_e}
    + \frac{f}{\kappa} = 0 \\
    \Rightarrow &
    f = \left(\kappa^{-1} + \sum_{e\in L\cup R} 
       \tilde k_e^{-1}
    \right)^{-1}
    \left( 
        \sum_{e \in L} \frac{F_e}{\tilde k_e} 
       - \sum_{e \in R} \frac{F_e}{\tilde k_e} 
    \right).
    \label{eq:f-value-new}
\end{aligned}\end{equation}

We can now evaluate the dissipation Eq.~\eqref{eq:diss} in the presence of the new edge $(m,n)$
by plugging in the relations \eqref{eq:change-of-F-new} and \eqref{eq:f-value-new},
\begin{equation}\begin{aligned}
    &D_{\rm loopy}\\ 
       &= \sum_{e \in T} \frac{F_e^2}{\tilde k_e}
       - \frac{\kappa}{ 1 + \sum_{e \in L\cup R} \frac{\kappa}{\tilde k_e}}
       \left( \sum_{e \in L} \frac{F_e}{\tilde k_e}
       - \sum_{e \in R} \frac{F_e}{\tilde k_e}
       \right)^2.
\end{aligned}\end{equation}

The average dissipation thus reads
\begin{equation}\begin{aligned}
    \langle D_{\rm loopy} \rangle
       &= \sum_{e \in T} \frac{\langle F_e^2 \rangle }{\tilde k_e}
       - \frac{\kappa}{ 1 + C_{m,n}\kappa } B_{m,n},
      \label{eq:D-loopy}  
\end{aligned}\end{equation}
where we introduced the abbreviations
\begin{equation}\begin{aligned}
     B_{m,n} &= \left\langle \left( 
              \sum_{e\in L} \frac{F_e}{\tilde{k}_e} -
              \sum_{e\in R} \frac{F_e}{\tilde{k}_e}
           \right)^2 \right\rangle \\
    C_{m,n} &=      \sum_{e \in L\cup R} \tilde{k}_e^{-1},\label{eq:abbreviations_dissipation}
\end{aligned}\end{equation}
which are functions only of the updated capacities $\tilde{k}_e$ along the sets of edges $L$ and $R$. Importantly, the resulting expression no longer contains the updated flows $\tilde{F}_e$, but only the flows in the tree network $F_e$, which are determined by Eq.~(\ref{eq:flows_treematrix}). This allows us to minimise the dissipation with respect to the updated capacities $\tilde{k}_e$ without having to take into account the interdependence of flows and capacities.\\ \\
\begin{figure*}[t!]
    \begin{center}
        \includegraphics[width=1.\textwidth]{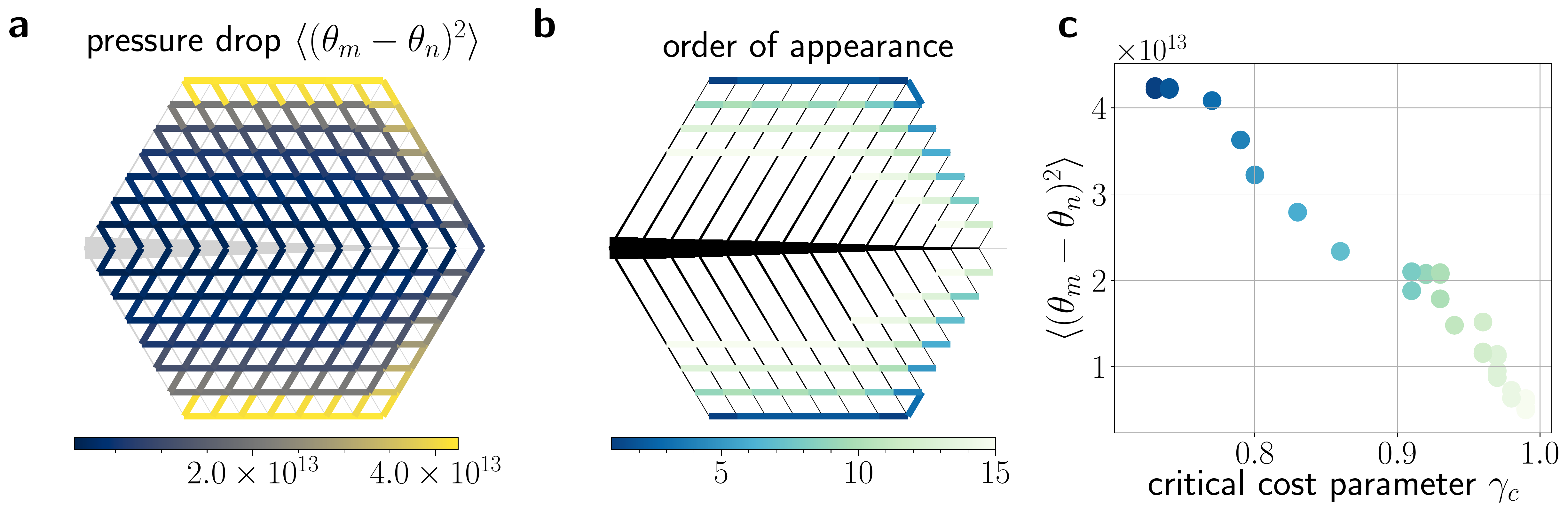}
    \end{center}
    \caption{\textbf{Average pressure drop predicts the order of appearance 
    of new loops as the cost parameter is varied.} \textbf{a} The average squared pressure drop $\langle (\theta_m-\theta_n)^2\rangle$ (colour code, cf. Eq.~\eqref{eq:pressure_drop}) calculated for the global tree optimum of dissipation allows to predict the location of the edges where loop formation first becomes beneficial. \textbf{b} Starting from a globally optimal tree network with cost parameter $\gamma = 0.4$, we slowly increase the cost parameter. We then determine in which order and at which critical value of the critical cost parameter $\gamma_c$ new edges appear closing a loop. \textbf{c} The pressure drop is strongly correlated with the critical cost parameter $\gamma_c$ where the given loop starts to form. Colour code corresponds to order of appearance of edges from dark (first) to light (last).
}
    \label{fig:loops_potential_drop}
\end{figure*}
Now that we have derived an explicit equation for the dissipation in near-tree networks, we will demonstrate how to minimise the resulting expression. For tree networks, the minima of the dissipation may be calculated explicitly using the method of Lagrange multipliers (see Supplementary Note 2). %
In contrast to that, we have to take into account an inequality constraint $\kappa \ge 0$ for near-tree networks as local minima may exist also at the boundaries of the domain. This can be done using the Karush-Kuhn-Tucker (KKT) conditions with the new Lagrange type function 
\begin{equation}\begin{aligned}
    \tilde{\LL}(\tilde{k}_e,\kappa) &= \sum_{e \in T} \frac{\langle F_e^2 \rangle }{\tilde k_e}- \frac{\kappa}{ 1 + C_{m,n} \kappa } B_{m,n}\\
    &-\tilde{\lambda} \left(K^\gamma - \sum_{e\in T} \tilde{k}^\gamma -\kappa^\gamma\right)-\mu \kappa\label{eq:KKT_Lagrangian},
\end{aligned}\end{equation}
where $\tilde{\lambda},\mu \in\mathbb{R}$ are KKT multipliers. The minimum is then determined by the KKT conditions (see Methods).

This approach results in explicit equations for the optimal edge capacities $\tilde k_e$, $\kappa$ in near-tree networks for which we could, however, not find a closed form solution for arbitrary networks and  values of $\gamma$ (see Supplementary Note 2). %
Still, we can make use of the resulting equations to gain insight into the process of loop formation. In particular, the KKT condition for the newly added edge $(m,n)$ with capacity $\kappa$ reads
\begin{equation}\begin{aligned}
   B_{m,n} =  (1 + C_{m,n} \kappa)^2 \kappa^{\gamma-1}\gamma\tilde \lambda \quad\lor\quad \kappa = 0\label{eq:complementary_slackness},
\end{aligned}\end{equation}
i.e. the capacity of the new edge either vanishes ($\kappa = 0$) or has the non-zero value given above. Importantly, we can obtain insights into the process of loop formation even without explicitly solving these equations.\\ \\

\noindent\textbf{How do loops emerge?}
We now illustrate how to make use of Eq.~\eqref{eq:complementary_slackness} to understand the process of loop formation. In particular, we rigorously demonstrate that loops form discontinuously as illustrated for the small tree network. Furthermore, we show that the tree remains a local minimum of the average dissipation even after the formation of a loop. We summarise these results in the following.%

\begin{theorem}[Tree remains KKT point]
\label{theo:local_minima}
Consider a linear flow network subject to the resource constraint with $\gamma\in(0,1)$. Then the following statements hold for the KKT points of the average dissipation $\langle D_{\rm loopy}\rangle$:
\begin{enumerate}
    \item There is always a KKT point at $\kappa = 0$, i.e. the tree is always a (potential) local minimum.
    \item The KKT point at $\kappa=0$ is isolated in the sense that we can find a real number $\varepsilon>0$ such that there are no other KKT points for $\kappa\in (0,\varepsilon)$.
\end{enumerate}
\end{theorem}
The proof makes use of the fact that we can find lower and upper bounds for the functions $B_{m,n}$, $\tilde{\lambda}$ and $C_{m,n}$ even without explicitly solving Eq.~\eqref{eq:complementary_slackness} and can be found in Supplementary Note 3. %
We note that the fact that the tree remains a local minimum is well-known for deterministic sources~\cite{banavar_topology_2000,bohn_structure_2007}.

We thus showed rigorously that for $\gamma<1$, KKT points that characterise a loopy network cannot emerge through a bifurcation of the local optimum describing a tree network network since the KKT point at $\kappa=0$ is isolated. Instead, new local minima of the dissipation generally emerge elsewhere and the transition to loopy networks is discontinuous. %
Having understood the mechanism of loop formation, we now proceed to analyse which edges will form the first loops.\\ \\
\noindent\textbf{Where do loops emerge first?}
We now study the location of the first loop that appears in the globally optimal network as the parameters of the model are varied. We start from the regime where the global optimum is a topological tree. 
Consider the expression for the average loopy dissipation $\langle D_{\rm loopy}\rangle$ to which a single edge $(m,n)$ of capacity $\kappa$ was added, as calculated in Eq.~(\ref{eq:D-loopy}).
We can find the location where loops form first by making the following approximation: assume that after the addition of the loop, the capacities of the edges $e$ along the shortest path from the source to the loop, $e \in R\cup L$, change only by a constant factor $c(\gamma,e)$, i.e. $\tilde k_e=c(\gamma,e)\, k_e$, whereas the other edges remain unchanged such that $c(\gamma,e)=1$ for these edges. Looking at Figure~\ref{fig:saddle_node_bifurcation} a, we can see that this is a reasonable assumption for the small network considered there. Note that the prefactor can be expected to be close to unity $c(\gamma,e)\approx 1$ even for edges $e\in L\cup R$ if we assume that the network is very large, because then the new edge will emerge with a very small capacity due to the resource constraint $\sum_e \tilde k_e^\gamma +\kappa^{\gamma} = K^{\gamma}$. 
With this approximation, the loopy dissipation reads
\begin{equation}\begin{aligned}
    \langle &D_{\rm loopy}\rangle  \\
    &\approx \sum_{e \in T} \frac{\langle F_e^2 \rangle }{c(\gamma,e)k_e}
       - \frac{\kappa}{ c(\gamma,e)^2\left(1 +\frac{C_{m,n}(k_e) \kappa}{c(\gamma,e)} \right)} B_{m,n}(k_e). 
\end{aligned}\end{equation}
Here, we defined the quantities $B_{m,n}(k_e)$ and $C_{m,n}(k_e)$ which we obtain from $B_{m,n}$ and $C_{m,n}$ (Eq.~\ref{eq:abbreviations_dissipation}) by replacing the updated capacity $\tilde{k}_e$ by the tree capacity $k_e$. The last expression can then be simplified considerably by making use of Eq.~(\ref{eq:kirchhoff}),
\begin{equation}\begin{aligned}
    B_{m,n}&(k_e) = \left\langle \left( \sum_{e\in L} \frac{F_e}{k_e} -
              \sum_{e\in R} \frac{F_e}{k_e}\right)^2 \right\rangle\\
              &= \left\langle (\theta_n-\theta_m)^2 \right\rangle.\label{eq:pressure_drop}
\end{aligned}\end{equation}
Thus, the emergence of loops is essentially governed by the potential drop across neighbouring vessels. Similar to how cracks in brittle materials form to relieve high elastic stresses, loop formation is determined by the relief of large pressure drops. Our explicit prediction is consistent with the idea that the reduction of pressure drops may have driven the evolution of leaf venation~\cite{Roth-Nebelsick2001}. From a developmental perspective, it connects to work explaining plant vein formation using models where mechanical stress relief is a crucial ingredient~\cite{Laguna2008,Couder2002,Bar-Sinai2016}. We confirm the `stress relief' by loop formation in terms of the potential drop by analysing the pressure drop before and after the formation of the loop in Supplementary Figure 1. %
\begin{figure*}[t!]
    \begin{center}
        \includegraphics[width=1.\textwidth]{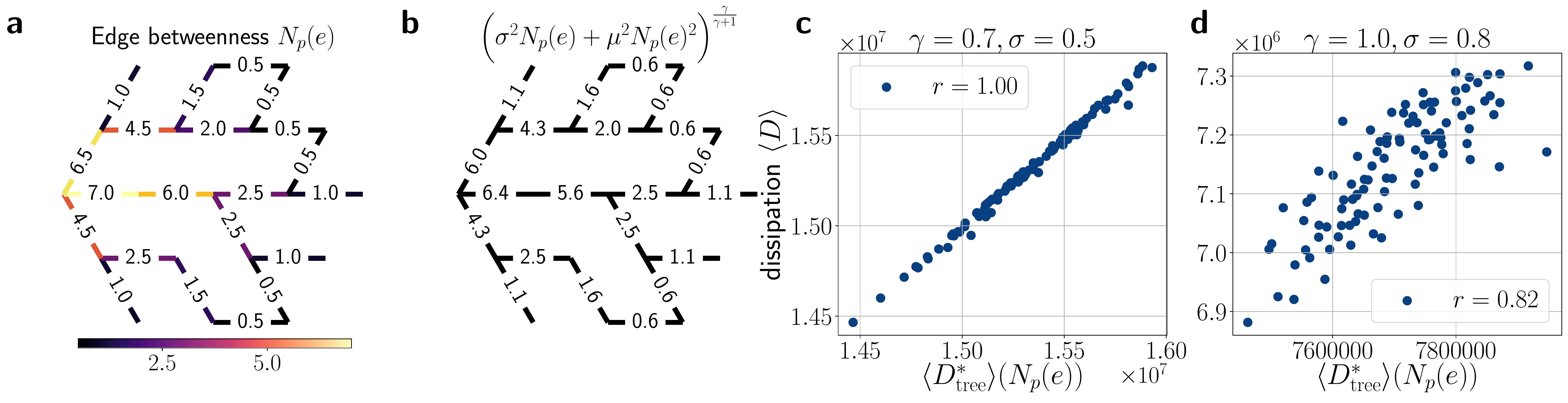}
    \end{center}
    \caption{\textbf{Edge betweenness centrality determines the network dissipation at local minima in the fluctuating sink model.} \textbf{a} Edge betweenness centrality $N_p(e)$ (numbers and colour code), determined with respect to a single source on the left, is closely related to the network dissipation. \textbf{b} The contribution $\left( N_p(e)\cdot \sigma^2+N_p(e)^2\cdot\mu^2\right)^{\frac{\gamma}{\gamma +1}}$ of a single edge to the minimal network dissipation in a tree network $\langle D^*_{\rm tree} \rangle$ as given in Equation~(\ref{eq:edge_betweenness_norm}) may be used to estimate the actual network dissipation at minima. Parameters used here are given by $\sigma = 0.5$, $\mu = 1$ and $\gamma = 0.9$. \textbf{c} The tree estimate $\langle D^*_{\rm tree} \rangle$ correlates strongly with the actual network dissipation at local minima $\langle D\rangle$ with high cost $\gamma = 0.7$ and low fluctuations $\sigma = 0.5$ since on average only $\langle N_L\rangle = 1.44$ loops form for this set of parameters (Pearson correlation coefficient of $r = 1.0$). \textbf{d} Moving to networks containing many loops, $\langle N_L\rangle = 36.66$ on average, obtained by minimising the dissipation for lower cost $\gamma = 0.8$ and more fluctuations $\sigma = 1.0$, the tree estimate still strongly correlates with the dissipation at minima as measured by a Pearson correlation coefficient of $r = 0.82$. Results were obtained by applying the relaxation method 100 times for each set of parameters where the set of  potential edges $\mathcal{E}$ forms a triangular network with $N=169$ nodes as shown in Figure~\ref{fig:5node_network}~c. 
    }
    \label{fig:betweenness_dissipation}
\end{figure*}

We now study the predictions made using Eq.~\eqref{eq:pressure_drop} numerically (see Methods for details). Starting from an optimal tree network, we first calculate the pressure drop (Fig.~\ref{fig:loops_potential_drop} a). We then successively decrease the cost for new edges and monitor the order in which new loops form (Fig.~\ref{fig:loops_potential_drop} b). Again, the transition to loop formation is discontinuous, such that loops emerge with a non-zero capacity at a critical value of the cost parameter $\gamma_c$, which is highly correlated to the pressure drop (Fig.~\ref{fig:loops_potential_drop} c). We may thus predict the location and cost parameter where loops form based on the potential drop.\\ \\
\textbf{Edge betweenness determines network dissipation.}
As we have demonstrated, all trees are -- and remain -- locally optimal structures and loopy networks emerge via saddle-node bifurcations. Thus, there may be a multitude of different local minima for a given set of network parameters, so a natural question that arises is the following: How can we determine which of the local minima have less dissipation than others and how can we find an order of different topologies, e.g. to find the topology that globally minimises the dissipation? Remarkably, we can trace back the answer to a purely topological property: the edge betweenness centrality.

We start by simplifying the locally optimal dissipation of the tree networks. In Eq.~(\ref{eq:flows_treematrix}), we expressed the flows $F_e$ in a tree network using the tree matrix
 $\mathcal{T}$. If we plug this expression into the self-consistency equation for the capacities Eq.~(\ref{eq:ke_corson}), set the overall available capacity to $K=1$, and plug everything into the dissipation Eq.~(\ref{eq:diss}) we arrive at the locally optimal dissipation in tree networks
\begin{equation}
\langle D^*_{\textrm tree} \rangle=\left[\sum_{e=1}^{N-1}\left( \sum_{i,j=2}^N \mathcal{T}_{e,j-1} \mathcal{T}_{e,i-1} \langle P_i P_j \rangle\right)^{\frac{\gamma}{\gamma +1}}\right]^{\frac{\gamma +1}{\gamma}}.
\end{equation}
Importantly, the entries appearing in this expression only depend on the mixed moments of the sinks and their second moments. Since the sinks are i.i.d. Gaussian random variables, these moments are identical for different sinks and are given by
\begin{equation}\begin{aligned}
    \langle P_i^2\rangle &= \mu^2 +\sigma^2,\quad i>1\\
    \langle P_iP_j\rangle &= \mu^2,~\qquad\quad i,j>1.
\end{aligned}\end{equation}
Thus, the sum runs over identical entries and we can calculate the dissipation as 
\begin{equation}\begin{aligned}
\langle D^*_{\rm tree} \rangle = \left[\sum_{e=1}^{N-1}\left(  N_p(e)\cdot \sigma^2+N_p(e)^2\cdot\mu^2\right)^{\frac{\gamma}{\gamma +1}}\right]^{\frac{\gamma +1}{\gamma}}\label{eq:edge_betweenness_norm}.
\end{aligned}\end{equation}
Here, $N_p(e)$ is the sum over the column of the tree matrix $\mathcal{T}$ that corresponds to edge $e$. In fact, $N_p(e)$ has the following interpretation: it is the {number of paths from the source $s$ to any other node $v$ that go through the edge $e$} and may thus be identified as a measure of shortest-path edge betweenness~\cite{anthonisse_rush_1971,newman_finding_2004} (see Methods). %
What can we learn from this analysis for loopy networks?

To estimate the contribution of a single edge to the overall network dissipation in a loopy network, we first calculate its edge betweenness (Fig.~\ref{fig:betweenness_dissipation} a) and, based on this, the contribution it would have to the dissipation in a tree network (Fig.~\ref{fig:betweenness_dissipation} b). Adding up the resulting expressions, we arrive at the tree estimate of the dissipation in a loopy network $\langle D^*_{\rm tree} \rangle (N_p(e))$. For near-tree networks, the correlation between the estimate and the actual dissipation at local minima is almost perfect as predicted by Eq.~\eqref{eq:edge_betweenness_norm} (Fig.~\ref{fig:betweenness_dissipation} c). Increasing the number of loops by tuning the noise parameter $\sigma$ and the cost parameter $\gamma$, edge betweenness and dissipation remain correlated even when there is a significant number of loops present in the network (Fig.~\ref{fig:betweenness_dissipation} d). Thus, we can still understand the minimal dissipation in loopy networks based on this topological measure.
 
We further discuss the possibility of characterising the global tree minima of the network dissipation in Supplementary Note 6.\\ \\
\section{Discussion}
In summary, we demonstrated that the transition to loop formation in optimal supply networks is discontinuous throughout different models and parameters. We explored this discontinuity in detail for a small example network, and rigorously proved that the discontinuous nature of the transition persists for larger networks as well. We showed that loops emerge through a saddle-node bifurcation, explaining the discontinuous transition.

Our results shed light on recent advances in the study of optimal supply networks. While the emergence of loops through fluctuations or damage was discovered recently~\cite{katifori2010damage,corson2010fluctuations}, the theoretical nature of this transition was until now not well understood. Here, we closed this gap by analysing the nature of the transition to loop formation in more detail. In particular, we obtained a measure of network stress that allowed us to predict the location and parameters where loops start to form. This opens a new pathway to the understanding of loop formation in natural networks such as leaves~\cite{katifori_quantifying_2012}.

Our results offer a new understanding of the interplay between the structure and function of supply networks. By unveiling the relationship between the network's topological edge betweenness and its average dissipation, we established a new link between form and function of networks. These results may aid in the understanding and design of {globally} optimal network structures such as biological vasculature, electrical grids, or neural networks. Our explicit prediction is consistent with the idea that the reduction of pressure drop variations may have been a factor in the evolution of leaf venation~\cite{Roth-Nebelsick2001}. More generally, we show that globally optimal network structures may be obtained by following simple local rules for adding new links, in contrast to previous work based on pruning an existing network~\cite{hu_adaptation_2013,ronellenfitsch_global_2016}.

Let us finally discuss how our results derived for linear flow models relate to other types of networks and systems. The starting point of our analysis was the fundamental trade-off between cost and resilience, which determines the optimal structure of a network, and which extends far beyond the theory of supply networks. Resilience requires additional capacity or links which can take over the load in the presence of failures or fluctuations -- but these are generally costly. From a practical view of network design, the fundamental question is thus: Where and how should new connections be added that increase resilience in an optimal way? 

Firstly, we discuss the question {where} new links should be added. A large body of literature in network science approaches aspects of resilience from the viewpoint of percolation theory. The fundamental question in this purely structural treatment is: Given a network, how many nodes or links may fail before the network gets disconnected? It has been shown that a decisive quantity to assess the resilience to random failures is given by the ratio of the second and first moment of the degree distribution, $\langle k^2 \rangle/\langle k^1 \rangle$. \cite{callaway2000network,newman2018networks}. These fundamental results were then used to optimise network resilience with respect to both random failures and targeted attacks \cite{shargel2003optimization,paul2004optimization}. In the case of failures, it is beneficial to add links between nodes which already have a high degree to effectively increase $\langle k^2 \rangle$. This result might appear very different to the findings of the present paper at first glance, but there are in fact common underlying principles. In supply network models, new links should be added where they will potentially attract a high flow. In percolation type models, new links should be added where they will potentially attract a high betweenness -- a quantity which can also be interpreted as a flow \cite{newman2005measure,anthonisse_rush_1971}. As a result, one should pick nodes whose characteristic quantity, either potential $\theta_n$ or degree $k$, differs from their surrounding. 

Secondly, we consider the question {how} new links should be added. A main finding of our work is that new links emerge in a discontinuous way with a finite non-zero capacity. That is, to be beneficial for the network, new links must have a certain minimum connection strength. This result has no direct equivalence in percolation approaches to network resilience, since the vast majority of studies in this field considers unweighted networks only. However, there is a strong interest in network formation processes, which induce discontinuities in macroscopic connectivity of the network -- including competitive percolation models \cite{achlioptas2009explosive,souza2015anomalous}, as well as transportation network models \cite{schroder2018hysteretic}.  In the context of network resilience, it has been shown that interdependencies and cascade effects can make the percolation transition discontinuous \cite{buldyrev2010catastrophic}.\\ \\
\section*{Methods}
\textbf{Numerical simulation of loop formation.}
When analysing the transition to loop formation such as in Figure~\ref{fig:discontinuous_transition_larger_network} and Figure~\ref{fig:loops_potential_drop}, we start from an optimal tree network $T$ for given parameters $\mu,\sigma$ and $\gamma$ corresponding to the dissipation minimum shown in Figure~\ref{fig:5node_network}c. As a next step, we add all non-tree edges from the underlying triangular grid with a very small capacity that corresponds to $1\%$ of the minimum capacity in the optimal tree, $k_f=0.01\cdot\operatorname{min}_{e\in T}(k_e^*)$, and then renormalise all capacities to make sure the resource constraint~\eqref{eq:capacity-constraint} holds. Finally, we then apply the iterative method described in Ref.~\cite{corson2010fluctuations} to let the new topology relax to a local minimum. If this minimum contains loops despite having started very close to the (global) tree minimum, and if its dissipation is lower than the one of the tree, we conclude that the given loopy topology is favourable.

To analyse the predictive power of the pressure drop in Fig.~\ref{fig:loops_potential_drop}, we initially consider a large optimised tree network with $N=169$ nodes and cost parameter $\gamma = 0.4$ for which we calculate the pressure drop (Fig.~\ref{fig:loops_potential_drop} a) and then increase the cost parameter from $\gamma=0.7$ to $\gamma = 0.99$, reoptimising the network for each value of gamma. Using the procedure outlined above, we compare the predicted positions of the first loops as indicated by the initial pressure drop $\langle (\theta_m-\theta_n)^2\rangle$ with the actual order in which they appear (Fig.~\ref{fig:loops_potential_drop} b). \\ \\
\textbf{Evaluating edge betweeenness}
In Eq.~(\ref{eq:edge_betweenness_norm}), we derived an alternative expression for the network dissipation at local minima that is based on the edge betweenness $N_p(e)$. The edge betweenness is defined as~\cite{anthonisse_rush_1971,newman_finding_2004,newman2018networks}
\begin{equation}\begin{aligned}
    N_p(e) = \sum_{t\in V}\frac{\sigma(s,t|e)}{\sigma(s,t)}.
    \label{eq:edge_betweenness}
\end{aligned}\end{equation}
Here, $\sigma(s,t)$ is the number of shortest paths from node $s$ to node $t$ and $\sigma(s,t|e)$ is the number of these shortest paths that contain the edge $e$. In the given setting, we consider this measure with respect to a single source $s$ that is identified as the source node of the network. Furthermore, when analysing tree networks, there is only one path from the source to every node $\sigma(s,t) = 1$ and thus $\sigma(s,t|e) =1 ~ \lor  = 0$.

In the main text, the edge betweenness is calculated using a method implemented in \textsc{python}'s \textsc{networkx} library~\cite{SciPyProceedings_11,brandes2001algorithm,brandes_variants_2008}.\\ \\
\textbf{Finding minima of a function with inequality constraints using KKT conditions.} Consider the function $D(\bm{k})$ of some real vector $\bm{k}=(k_1,...,k_N)^\top\in\mathbb{R}^N$ that is subject to the equality constraint $h(\bm{k}) = 0$ and the inequality constraint $g(\bm{k})\leq 0$ which we assume to be described by differentiable, real-valued functions $g,f:\mathbb{R}^N\rightarrow \mathbb{R}$. To identify potential maxima or minima of the function subject to the constraints, we can make use of the KKT conditions. To this end, we consider the Lagrange type function
\begin{equation}\begin{aligned}
    \tilde{\LL}(\bm{k}) &= D(\bm{k})+\tilde{\lambda} h(\bm{k}) +\mu g(\bm{k}),
\end{aligned}\end{equation}
where $\tilde{\lambda},\mu\in\mathbb{R}$ are called KKT multipliers. Then the following conditions, the KKT conditions, are necessary condition for a point $\bm{k}^*$ being a minimum of $D(\bm{k})$ ~\cite{boyd_convex_2004,kuhn1951}
\begin{equation}\begin{aligned}
    &\frac{\partial \tilde{\LL}}{\partial k_i^*}\stackrel{!}{=} 0,\quad\forall i \in \{1,...N\}, \\
    &f(\bm{k}^*)=0,\\
    &g(\bm{k}^*)\leq 0,  \\
    &\mu\geq 0,\\
    &\mu g(\bm{k}^*) = 0.
\end{aligned}\end{equation}
This formulation may be used to find out whether adding a single loop to a tree network may reduce its dissipation.
\\ \\
\normalsize\textbf{Data availability.} Photographs of leaf venation networks in Figure~\ref{fig:real_world_networks} are available upon request. The topology of the Scandinavian power grid  has been extracted from the open European energy system model PyPSA-Eur~\cite{horsch_2018}, which is fully available online at \url{https://doi.org/10.5281/zenodo.3886532}. Distribution grid in panel c was extracted from Ref.~\cite{Kersting_2001}.
\\ \\
\textbf{Code availability.} 
Computer code will be made available at \url{https://github.com/FNKaiser/Optimal_Supply_Networks} upon publication.
\\
\\

\bibliographystyle{naturemag}
\bibliography{network}
\bigskip{}
\noindent{\textbf{Acknowledgements}}\\
\normalsize We thank Eleni Katifori for valuable discussions. We gratefully acknowledge support from the Federal Ministry of Education and Research (BMBF grant no. 03EK3055B D.W.), the Helmholtz Association (via the joint initiative "Energy System 2050 -- a Contribution of the Research Field Energy" and grant no. VH-NG-1025 to D.W.).\\ \\
\noindent\textbf{Author Contributions}\\
\normalsize D.W. conceived research and acquired funding. F.K. and D.W. designed research. F.K. and H.R. carried out numerical simulations. F.K. evaluated the results and designed all figures. All authors contributed to discussing the results and writing the manuscript.\\
\\
 \normalsize\textbf{Competing Interests} \\
The Authors declare no Competing Financial or Non-Financial Interests.\\ \\
\textbf{Additional Information}\\ 
\textbf{Supplementary Information} is available for this paper.\\ \\
   \textbf{Correspondence and requests for materials} should be addressed to Dirk Witthaut~(email: d.witthaut@fz-juelich.de).
%\clearpage
%\newpage
%  \includepdf[pages=1-]{Optimal_Networks_SI.pdf} 
\end{document}